\documentclass[letterpaper, 10 pt, conference]{ieeeconf}  % Comment this line out if you need a4paper
\IEEEoverridecommandlockouts 
\pdfoutput=1
\usepackage{amsmath,amssymb,color,cite}%,showkeys}
\usepackage{graphicx}

\usepackage{latexsym}
\usepackage{algorithm}
\usepackage{algpseudocode}
\usepackage{verbatim}
\usepackage{cite}
\usepackage[english]{babel}
\usepackage{balance}

\let\labelindent\relax
\usepackage{amsmath,amsthm}
\usepackage{enumitem}

\usepackage{xcolor}

\usepackage{pgf}
\usepackage{pgfplots,tikz}
\usetikzlibrary{tikzmark}
\usepackage{arydshln}

\usepackage{amsfonts}
\usepackage{amssymb}
\usepackage{amsbsy}
\usepackage{bm}
\usepackage{mathtools}
\usepackage[font=scriptsize]{subcaption}
\usepackage[font=scriptsize]{caption}
\theoremstyle{definition}
\newtheorem{definition}{Definition}

\theoremstyle{plain}
\newtheorem{assumption}{Assumption}
\newtheorem{lemma}{Lemma}
\newtheorem{theorem}{Theorem}

\newtheorem{Remark}{Remark}[section]
\newcommand{\bRemark}{ \begin{Remark} \rm }
\newcommand{\eRemark}{ \end{Remark}    }

\newcommand{\Scal}{\mathcal{S}}
\newcommand{\Mcal}{\mathcal{M}}
\newcommand{\Acal}{\mathcal{A}}
\newcommand{\Hcal}{\mathcal{H}}
\newcommand{\EE}{\mathbb{E}}
\newcommand{\PP}{\mathbb{P}}
\newcommand{\pistar}{{\pi^*}}
\newcommand{\Vstar}{V^*}
\newcommand{\Pitilde}{\widetilde{\Pi}}
\newcommand{\hthresh}{{\bar{h}}}
\newcommand{\pithresh}{\pi_{t,\bar{h}}}
\DeclareMathOperator*{\argmin}{arg\,min}

\title{\LARGE \bf Tiered Service Architecture for Remote Patient Monitoring 
}

\author{Siddharth Chandak$^{1}$, Isha Thapa$^{2}$, Nicholas Bambos$^{1,2}$ and David Scheinker$^{2,3}$\\ 
$^{1}$ Department of Electrical Engineering, Stanford University, USA.\\
$^{2}$ Department of Management Science \& Engineering, Stanford University, USA.\\
$^{3}$ School of Medicine, Stanford University, USA.\\
{ \{chandaks, ishadt, bambos, dscheink\}@stanford.edu}
}

\begin{document}
\maketitle

\begin{abstract}
We develop a remote patient monitoring (RPM) service architecture, which has two tiers of monitoring: ordinary and intensive. The patient's health state improves or worsens in each time period according to certain probabilities, which depend on the monitoring tier. The patient incurs a ``loss of quality of life" cost or an ``invasiveness" cost, which is higher under intensive monitoring than under ordinary. On the other hand, their health improves faster under intensive monitoring than under ordinary. In each period, the service decides which monitoring tier to use, based on the health of the patient. We investigate the optimal policy for making that choice by formulating the problem using dynamic programming. We first provide analytic conditions for selecting ordinary vs intensive monitoring in the asymptotic regime where the number of health states is large. In the general case, we investigate the optimal policy numerically. We  observe a threshold behavior, that is, when the patient's health drops below a certain threshold the service switches them to intensive monitoring, while ordinary monitoring is used during adequately good health states of the patient. The modeling and analysis provides a general framework for managing RPM services for various health conditions with medically/clinically defined system parameters.

\end{abstract}

% \begin{IEEEkeywords}
% digital health, patient monitoring over the internet, health care management
% \end{IEEEkeywords}

\section{Introduction}

Remote Patient Monitoring (RPM) is increasingly receiving attention as a method for monitoring patients with certain medical conditions in their normal living/working environments to increase their quality of life and the level of delivered health care \cite{Farias, Malasinghe, Zinzuwadia}. This is becoming feasible via advancements in wearable medical devices, for example, wearable glucose monitors \cite{Lee}, smart watches with vital sign monitoring capabilities (e.g., heart rate, ECG, pulse oximetry) \cite{Masoumian, BarDavid}, and other such sensors. Further, such devices are increasingly networked and can transmit and receive data over the Internet and act as edge devices in communication with computation servers in the Cloud.

Studies have shown the effectiveness of RPM for various medical conditions. For example, continuous glucose monitoring has been shown to improve glycemic control in patients with diabetes \cite{Maiorino,prahalad}. Smart watches have been effectively used to monitor stress, movement disorders, sleep patterns, blood pressure, heart disease, and COVID-19 \cite{Masoumian}. Other RPM devices have also been used to manage and track cardiac conditions, such as heart failure, arrhythmia, and hypertension \cite{Zinzuwadia}. These studies highlight the potential of RPM to improve patient outcomes and quality of life by allowing for timely interventions and personalized care. 

However, the question remains of how intensively to monitor patients. Intensive monitoring schemes could range from remotely collecting more data on the patient health state and administering more medical intervention remotely (e.g., alerting the patient to increase medication dosage) to calling the patient into an urgent care facility.

While aggressive monitoring may provide more comprehensive data, it can be resource intensive, draining the wearable device battery \cite{BarDavid} faster and requiring the clinicians to review more RPM data. From the patient's perspective, under intensive monitoring, the patient experiences a higher ``loss of quality of life'' cost or ``invasiveness'' cost, since intensive monitoring would normally be more invasive to their personal lifestyle and can result in treatment fatigue \cite{heckman}. In contrast, intensive monitoring (and correspondingly elevated medical intervention) would enable early detection and intervention for adverse events, and hence the patient's health is expected to improve faster than under ordinary one. Therefore, to account for this trade-off between the invasiveness cost and the possibility of an early intervention, there is an inherent need for a systematic approach to determining the appropriate level of monitoring, based on the patient's health state.

In this paper we develop a RPM service architecture, where the patient is placed under less or more intensive levels or tiers of monitoring, based on their health state. We then study the optimal monitoring strategy for this model and how it varies with different parameters. One would intuitively expect that a patient would be placed under intensive one when their health state deteriorates; on the other hand, they would be returned to ordinary monitoring when their health state improves enough. The decision to switch from ordinary to intensive monitoring (and/or vice versa) requires a systematic analysis and depends heavily on the various parameters of the service in a rather subtle and complicated way, as analyzed in the following sections.  

Of course, the design of an RPM service for a specific medical condition is highly dependent on the specifics of that condition and requires specialized medical knowledge. The point of this paper is not to design a particular RPM service but to provide a general framework and systematic methodology for RPM services based on tunable parameters (e.g., health improvement or deterioration probabilities, monitoring options and invasiveness costs) so as to make justifiable monitoring choices. The parameters will have to be decided on and tuned by medical/clinical experts for condition-specific RPM services. While we work with a simplified model here, the intuition gained from the analysis can help clinicians take more informed monitoring decisions.

In section \ref{sec:model}, we develop the model of evolution of the patient health state under ordinary and intensive monitoring and demonstrate how it can be managed, using the methodology of dynamic programming. In section \ref{sec:analysis}, we provide some analytical results on the optimal management policy in the asymptotic regime of a large number of health states and provide conditions on the parameters for choosing ordinary vs intensive monitoring. In section \ref{sec:perfomance}, we numerically investigate the structure of optimal monitoring policies and demonstrate that they place the patient under intensive monitoring when the health state deteriorates below a certain threshold; otherwise they use ordinary monitoring. Finally, in section \ref{sec:conclusions}, we discuss some extensions. Appendix \ref{app:proofs} contains proofs for results presented in section \ref{sec:analysis}.

\section{The RPM Service Model} 
\label{sec:model}

Consider a patient who can be in a {\bf health state} $h_t\in\{0,1,2,3,...,H\}$ in each service time period $t\in\{0,1,2,3,...\}$. The RPM service places the patient in a monitoring/intervention state $m_t\in\Mcal=\{o,i\}$ in each time period $t$, abbreviated to {\bf monitoring state}, where $o$ denotes ordinary monitoring and $i$ intensive monitoring. Thus, one can view the 
monitoring-patient joint state $s_t\coloneqq(m_t, h_t)\in\Mcal\times\Hcal\eqqcolon\Scal$ as the system or service state at time $t$.

A higher patient health state $h\in\{0,1,2,3,...,H\}$ corresponds to the patient having better health. In particular, the lowest health state $0$ is {\bf critical} in the sense that, when the patient drifts into that state, they go beyond the scope of the current service model; at that point other emergency and/or more severe medical interventions are required, which are outside the scope of this service. Because of that, the states $(i,0)$ and $(o,0)$ are absorbing for the Markovian evolution of the health state, as explained below. Indeed, when the patient enters heath state $0$ under any monitoring state $i$ or $o$, the service evolution stops, as other medical measures/interventions are initiated.

As seen below, in defining costs incurred at the various states, we first take the patient's quality of life point of view. Under ordinary monitoring, the patient incurs a constant cost $C_o\ge0$ at any state $(o,h)$ with $h\in\{1,2,...,H\}$. Correspondingly, under intensive monitoring, the patient incurs a constant cost $C_i\ge0$ at any state $(i,h)$ with $h\in\{1,2,..,H\}$. These costs reflect the invasiveness loss for the patient. One may argue about the costs in more elaborate ways, for example, including patient risk factors and operational considerations of the service. For simplicity, we focus here on the invasiveness argument mentioned above. 

Of special interest is the critical health state $h=0$, where this model ceases to apply. On either state $(o,0)$ or $(i,0)$ where the patients health is critical and a cost $C_c$ is incurred. 

We finally define the transition costs $C_{o\to i}$ (and $C_{i\to o}$), which is associated with the service transitioning the patient from ordinary to intensive monitoring (and vice versa, respectively).

We model the system as a controlled Markov chain. Such models are commonly used for medical decision making \cite{steimle, alagoz}. We try to stay as simple as possible, yet still capture the essence of the problem and get insights into its solution. At the beginning of every time period $t$, the service takes the decision/action (control) to either keep the monitoring state the same (as in the previous time period) or switch it to the alternate monitoring state. Formally, the decision/action space is $\Acal=\{o,i\}$ and each (state, action) pair is associated with a cost given by the function $c:\Scal\times\Acal\mapsto\mathbb{R}^+$. The transition probabilities are given by $p(s'|s,a)$ where $s',s\in\Scal$ and $a\in\Acal$. The cost functions and transition probabilities are defined as follows.

\noindent\textbf{1. At health state $\mathbf{h=0}$:} 
\begin{quote}
    No action is taken with the the service ceasing operation. A cost of $C_c$ is incurred.
\end{quote}

\noindent \textbf{2. At health states $\mathbf{1\leq h\leq H}$}:
\begin{enumerate}[label=\alph*)]
\item {\em Ordinary Monitoring ($m=o$), no Switching ($a=o$):} 
Does not induce a monitoring change, and the system state transitions  as follows:
\begin{enumerate}[label=\roman*)]
\item 
$(o,h)\xrightarrow{a=o} (o,\min\{h+1,H\})$ with prob. $\lambda_o$
\item
$(o,h)\xrightarrow{a=o} (o,h-1)$ with prob. $\mu_o=1-\lambda_o$,
\end{enumerate}
and a cost $c\big((o,h),o\big)=C_o$ is incurred. Note that $\min\{h+1,H\}$ above is used to account for $(o,H)\to (o,H)$ with prob. $\lambda_o$ since $H$ is the highest health state. 

\item {\em Ordinary Monitoring ($m=o$) with Switching ($a=i$):}
Induces a switch to intensive monitoring $m=i$, and the system state transitions as follows:
\begin{enumerate}[label=\roman*)]
\item 
$(o,h)\xrightarrow{a=i} (i,\min\{h+1,H\})$ with prob. $\lambda_i$
\item
$(o,h) \xrightarrow{a=i} (i,h-1)$ with prob. $\mu_i=1-\lambda_i$,
\end{enumerate}
and a cost $c\big((o,h),i\big)=C_{o\to i}+C_i$ is incurred.

\item {\em Intensive Monitoring ($m=i$), no Switching ($a=i$):}
Does not induce a monitoring change, and the system state transitions as follows:
\begin{enumerate}[label=\roman*)]
\item 
$(i,h)\xrightarrow{a=i}(i,\min\{h+1,H\})$ with prob. $\lambda_i$
\item
$(i,h)\xrightarrow{a=i}(i,h-1)$ with prob. $\mu_i=1-\lambda_i$,
\end{enumerate}
and a cost $c\big((i,h),i\big)=C_i$ is incurred.

\item {\em Intensive Monitoring ($m=i$) with Switching ($a=o$):}
Induces a switch to ordinary monitoring $m=o$, and the system state transitions as follows:
\begin{enumerate}[label=\roman*)]
\item 
$(i,h) \xrightarrow{a=o} (o,\max\{h+1,H\})$ with prob. $\lambda_o$
\item
$(i,h)\xrightarrow{a=o}(o,h-1)$ with prob. $\mu_o=1-\lambda_o$,
\end{enumerate}
and a cost  $c\big((i,h),o\big)=C_{i\to o}+C_o$ is incurred.
\end{enumerate}

We can easily incorporate health state dependent costs and transition probabilities, but for simplicity we assume constant ones here. We make the following \textit{natural} assumptions.
\begin{assumption}
\begin{enumerate}[label=\alph*)]
    \item The transition probabilities satisfy: $\lambda_i\geq\lambda_o$.
    \item The costs satisfy: $0\leq C_o\leq C_i\leq C_c$.
\end{enumerate}
\end{assumption}
The first assumption 1.a) intuitively states that the patient's health improves faster under intensive monitoring, rather than under ordinary. Regarding assumption 1.b), it is naturally expected  that  $C_o\le C_i$, as the patient's ``annoyance" is higher under intensive monitoring/intervention than under ordinary. Further, given the severity of entering the critical state $h=0$, it is naturally expected that $C_i\le C_c$, and practically $C_c$ is expected to be {\em much larger} than $C_i$.

\subsection{Optimal Monitoring Control}

We study this problem under the {\em discounted cost} setting of the dynamic programming methodology \cite{Bertsekas}, hence, costs incurred $t$ time periods into the future (with respect to present) are discounted by a factor of $\gamma^t$ with $0<\gamma<1$. Starting from state $s_0=s\in\Mcal\times\Hcal$, the total expected (discounted) cost to be incurred is
\begin{equation*}
\EE\Big[\sum_{t=0}^{T-1} \gamma^tc(s_t,a_t)+\gamma^TC_c \Big | \ s_o=s\Big]    
\end{equation*}
when control action $a_t$ is taken by the service, at cost $c(s_t,a_t)$ introduced above, when its state is $s_t=(m_t,h_t)$ at time $t$, until the patient enters the critical state $h=0$ at time $T$ and the service ceases operation. Hence, $T$ is the time the patient spends in the service, and at time $T$ the critical cost $C_c$ is incurred, however, discounted to $\gamma^TC_C$. Thus, discounting by $\gamma$ implicitly reflects the patient's desire to stay longer in service, hence, incur the critical cost further in future and discounted to $\gamma^T C_c$.

A (stationary) monitoring policy $\pi(s)$ is a rule mapping each state $s=(m,h)\in\Mcal\times\Hcal$ to a control action $a\in\Acal=\{o,i\}$ to be taken at that state. The value function $V_\pi(s)$ of a policy $\pi(s)$ is the total expected (discounted) cost the system will incur until reaching the critical state $0$ and stop, when it starts from state $s$ at time $t=0$. That is,
$$
V_\pi(s)=\EE\left[\sum_{t=0}^{T-1}
\gamma^t c\Big(s_t,\pi(s_t)\Big)
+\gamma^TC_c\ \Big |\ s_0=s\right]
$$
and satisfies the dynamic programming equation \cite{Bertsekas}.
$$
V_\pi(s)=c\Big(s,\pi(s)\Big)+
\gamma
\sum_{s'\in\Scal} \PP\Big(s'\mid s,\pi(s)\Big) V_\pi(s'),
$$
for all $s\in\Scal=\Mcal\times\Hcal$, given the Markovian evolution dynamics of the system, specified by the state transition probabilities defined above.

The goal is to find an optimal policy $\pi^*$ which minimizes $V_\pi$ over all policies $\pi$, i.e., $V_{\pi^*}(s) \le V_\pi(s)$ for every $s\in\Scal$ over all policies $\pi$. For simplicity, we define $V^*(s)\coloneqq V_{\pi^*}(s)$ which satisfies the following dynamic programming equation \cite{Bertsekas}.
$$
\Vstar(s)=\min_{a\in\{o,i\}} \Bigg\{c(s,a)+\gamma
\sum_{s'\in\Scal}\PP\Big(s'|s,a\Big)\Vstar(s')\Bigg\},
$$
and can be solved numerically to yield the optimal policy $\pi^*(s)=\pi^*(m,h)$, that is, what optimal decision to take when the patient is in health state $h$ under monitoring $m$. For the state transition probabilities and costs defined before, this dynamic programming equation unfolds into:

\begin{enumerate}
[align=left, 
leftmargin=0pt, 
labelindent=\parindent, listparindent=\parindent, 
labelwidth=0pt, 
itemindent=!,
label=(\roman*)]
\item 
For health states $1\leq h\leq H-1$
\begin{align*}
    &\Vstar(i,h)\nonumber =\min\bigg\{C_i+\gamma\Big[\lambda_i\Vstar(i,h{+}1)+\mu_i\Vstar(i,h{-}1)\Big],\nonumber\\
    &\;\;\;\;\;\;\;\;C_{i\to o}+C_o+\gamma\Big[\lambda_o\Vstar(o,h{+}1)+\mu_o\Vstar(o,h{-}1)\Big]\bigg\},
\end{align*}
\begin{align*}
    &\Vstar(o,h)\nonumber=\min\bigg\{C_o+\gamma\Big[\lambda_o\Vstar(o,h{+}1)+\mu_o\Vstar(o,h{-}1)\Big],\nonumber\\
    &\;\;\;\;\;\;\;\;C_{o\to i}+C_i+\gamma\Big[\lambda_i\Vstar(i,h{+}1)+\mu_i\Vstar(i,h{-}1)\Big]\bigg\}.
\end{align*}

\item At health state $H$, 
\begin{align*}
    &\Vstar(i,H)\nonumber=\min\bigg\{C_i+\gamma\Big[\lambda_i\Vstar(i,H)+\mu_i\Vstar(i,H{-}1)\Big],\nonumber\\
    &\;\;\;\;\;\;\;\;C_{i\to o}+C_o+\gamma\Big[\lambda_o\Vstar(o,H)+\mu_o\Vstar(o,H{-}1)\Big]\bigg\},
\end{align*}
\begin{align*}
    &\Vstar(o,H)\nonumber=\min\bigg\{C_o+\gamma\Big[\lambda_o\Vstar(o,H)+\mu_o\Vstar(o,H{-}1)\Big],\nonumber\\
    &\;\;\;\;\;\;\;\;C_{o\to i}+C_i+\gamma\Big[\lambda_i\Vstar(i,H)+\mu_i\Vstar(i,H{-}1)\Big]\bigg\}.
\end{align*}
\item 
At the critical health state $h=0$, 
\begin{equation*}
\Vstar(i,0)=\Vstar(o,0)=C_c.
\end{equation*}
\end{enumerate}

Note that in every $\min\{\cdot,\cdot\}$ above, the first term corresponds to the control keeping the existing monitoring state, while the second corresponds to the control switching to the alternate monitoring state and incurring the switching cost.

\subsection{Simplified RPM Service}

In order to reduce the number of parameters for tractability of the analysis, we make the following simplification.

\begin{definition} {\bf Simplified RPM Service:}.
\begin{enumerate}[label=\alph*)]
\item 
The cost for ordinary monitoring is set to zero: $C_o=0$.
\item 
The switching costs are set to zero: $C_{i\to o}=C_{o\to i}=0$.
\end{enumerate}
Assumption 1 is still satisfied.
\end{definition}

% The first assumption is that the cost of invasiveness under ordinary monitoring is zero. As can be expected, this assumption does not affect the results significantly. Under the second property, there is no cost for the patients to transfer between different monitoring services. While non-zero transition costs allow for a richer model, this property can be justified in several applications where it is easier to implement intensive monitoring.

Given the limited space of this short paper, we work with the simplified RPM service below (analysis and numerical results), where interesting insights emerge. The zero cost of invasiveness under ordinary monitoring has no significant impact on the results, and is merely a technical assumption to make our analysis easier. The non-zero transition cost is true in several applications where the intensive monitoring just involves a higher rate of collecting data about the patient's health. We briefly comment on how non-zero transition costs affect our results in Section \ref{sec:conclusions}.

The simplified RPM service is illustrated in Figure \ref{fig:model} below.

\begin{figure}[h!]
\centering
\includegraphics[width=0.8\linewidth]{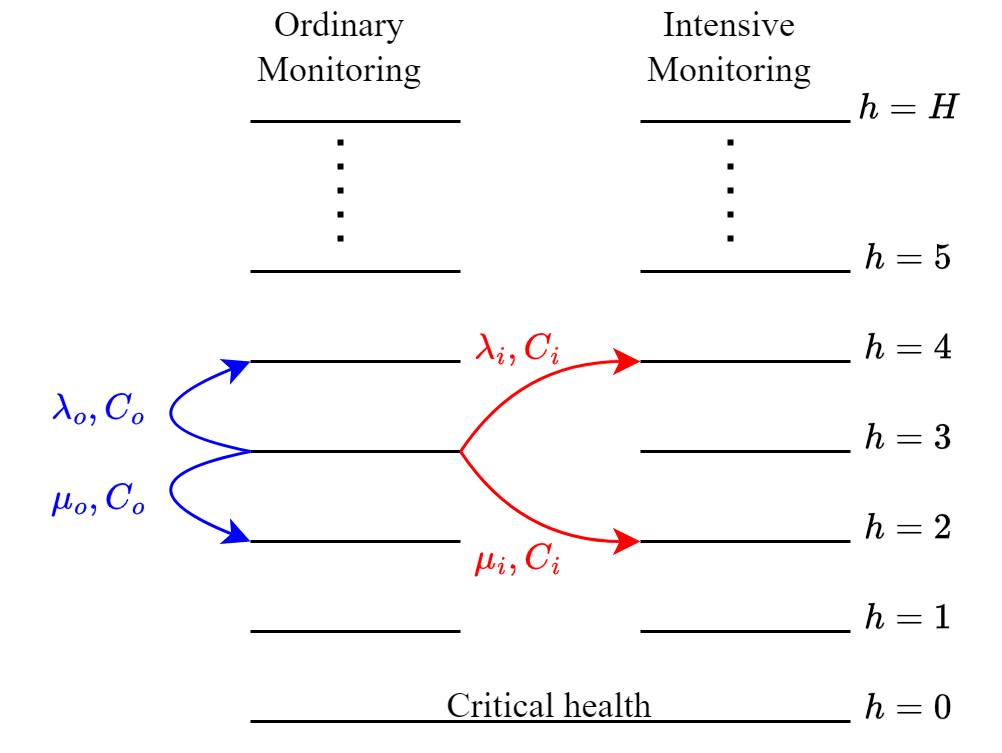}
\caption{The Simplified RPM Service. The blue and red arrows represent the possible transitions from state $(o,3)$ for actions $o$ and $i$ respectively. The arrows are labeled with probability of transition and cost incurred.}
  \label{fig:model}
\end{figure}

The dynamic programming equations for $V^*$ given above reduce in this case to the following:
\begin{enumerate}
[align=left, 
leftmargin=0pt, 
labelindent=\parindent, listparindent=\parindent, 
labelwidth=0pt, 
itemindent=!,
label=(\roman*)]
\item 
For health states $1\leq h\leq H-1$,
\begin{multline}\label{dp-general}
\hskip -0.3 cm 
\Vstar(i,h)=\Vstar(o,h) \\
\hskip -0.3cm
=\min\Bigg\{ C_i + \gamma \Big[ \lambda_i\Vstar(i,h+1) + \mu_i \Vstar(i,h-1) \Big], \\
\hskip 1.35cm
C_o + \gamma \Big[ \lambda_o \Vstar(o,h+1) + \mu_o\Vstar(o,h-1)\Big] \Bigg\}
\end{multline}
\item 
At health state $H$, 
\begin{multline} \label{dp-H}
\Vstar(o,H)=\Vstar(i,H) \\
=\min\Bigg\{ C_i + \gamma \Big[ \lambda_i \Vstar(i,H)+\mu_i\Vstar(i,H-1) \Big], \\
C_o + \gamma \Big[ \lambda_o \Vstar(o,H) + \mu_o \Vstar(o,H-1) \Big] \Bigg\}
\end{multline}
\item 
At the critical health state $h=0$, 
\begin{equation} \label{dp-final}
\Vstar(i,0)=\Vstar(o,0)=C_c,
\end{equation}
\end{enumerate}
Note that, in the absence of switching costs (that is, $C_{o\to i}=C_{i\to o}=0$), for any health state $h$ the $\Vstar$ is the same under both ordinary and intensive monitoring.

\section{Asymptotic Analysis}\label{sec:analysis}

The optimal policy $\pistar$ and value function $\Vstar$ can be computed numerically (in general) from the dynamic programming equations \eqref{dp-general}-\eqref{dp-final} for any given set of system parameters. But to develop intuition and characterize the optimal policy, in this section, we analyse $\pistar$ in the asymptotic regime of a large number of health states $H\gg1$, i.e., a dynamic range of health states $H\to\infty$. This allows for analytic tractability of the optimal policy and closed-form conditions on our policies of interest. We make the following assumptions in this section.

\begin{assumption}
{\bf Large $H$ Asymptotic Regime.}
\begin{enumerate}[label=\alph*)]
\item 
The number of health states is very large, i.e., the system operates in the asymptotic regime of $H\to\infty$.
\item 
Under ordinary monitoring the patient's health drifts downwards, i.e., the improvement probability is  $\lambda_o<0.5$ and the worsening one is $\mu_o=1-\lambda_o>0.5$
\end{enumerate}
\end{assumption}
Assumption 2.a) allows us to use tools from random walk analysis, as done in Lemma \ref{lemma:out}, making the analysis tractable. Assumption 2.b)  allows for the Markov chain to remain stable (positive recurrent) in the asymptotic regime. 

The results obtained in this asymptotic regime can be thought of as an approximation for the RPM service with finite $H$, when $H$ grows large. In the next section, we numerically demonstrate that this asymptotic approximation tracks the optimal policy and the value function for our RPM for a number of health states as low as $H=5$.

We define $\Pitilde$ as the the set of policies under which the service chooses the same action irrespective of the monitoring state, that is,
$$
\Pitilde=\{\pi\mid \pi(i,h)=\pi(o,h), \;\forall h\geq 1\}.
$$
Based on \eqref{dp-general} and \eqref{dp-H}, for all $h\geq 1$, we then have
\begin{align*}
    &\pistar(o,h)=\pistar(i,h)\nonumber\\
    &=\argmin_{\{i,o\}}\{C_i+\gamma\lambda_i\Vstar(i,h+1)+\gamma(1-\lambda_i)\Vstar(i,h-1),\nonumber\\
    &\;\;\;\;\;\;C_o+\gamma\lambda_o\Vstar(o,h+1)+\gamma(1-\lambda_o)\Vstar(o,h-1)\}
\end{align*}
This implies that $\pistar\in\Pitilde$ and we can restrict our attention to the set $\Pitilde$. For simplicity, we introduce notation $\Vstar(h)=\Vstar(o,h)=\Vstar(i,h)$ for $h\in\Hcal$ and similarly the notation $\pistar(h)$. 

% This also enables us to refer to the policy as the patient prefers to stay in ordinary monitoring at health state $h$ under policy if $\pistar(h)=o$ and similarly for intensive monitoring. 

We next define an important policy $\pi_o$ where the patient stays under ordinary monitoring at all health states,  i.e., $\pi_o(i,h)=\pi_o(o,h)=o$ for all $h\geq1$. Note that $\pi_o\in\Pitilde$ and we define the corresponding value function $V_o(h)$. Our first lemma gives the value function for this policy and presents an important property about the optimal policy.
\begin{lemma}\label{lemma:out}
    For the simplified RPM (Definition 1) and under Assumption 2, 
    \begin{enumerate}[label=\alph*)]
        \item The value function $V_o$ for the policy $\pi_o$ is given by:
        $$V_o(h)=\phi^hC_c,$$
        where $$\phi=\frac{1-\sqrt{1-4\lambda_o\mu_o\gamma^2}}{2\lambda_o\gamma}.$$
        Note that $\phi<1$ for $\gamma<1$.
        \item For any choice of parameters, there exists $h'$ such that, under the optimal policy, the patient prefers to stay in ordinary monitoring above health state $h'$, i.e., $\pistar(h)=o$, for all $h\geq h'$.
    \end{enumerate}
\end{lemma}
\begin{proof}
    See Appendix.\ \ref{app:proofs} for proofs.
\end{proof}
For the simplified RPM, the cost of invasiveness under ordinary monitoring is zero. Then $V_o(h)=\EE[\gamma^TC_c|h_0=h]$, where $T$ is the time taken to reach health state $0$, when started at health state $h$ and when the patient always stay under ordinary monitoring. $T$ here is precisely the hitting time of state $0$ for a random walk initiated at state $h$. The proof for this lemma then follows from the moment generating function of the hitting time for an $\infty$-state random walk.

\begin{figure*}[h!]
\centering
\begin{subfigure}[t]{0.32\textwidth}
  \centering
  \includegraphics[width=0.85\linewidth]{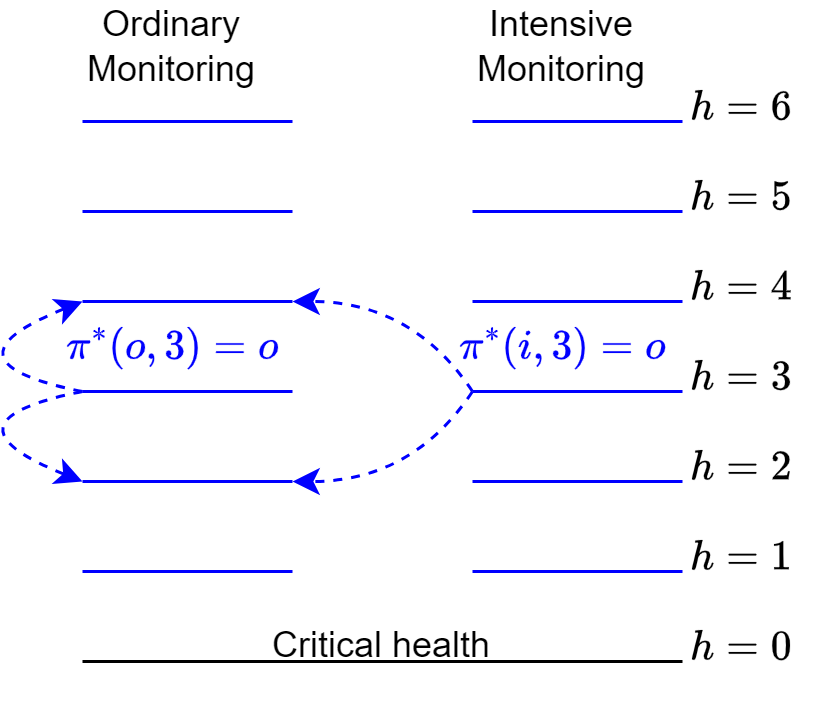}
  \caption{Optimal policy is $\pi_o$ (always in ordinary monitoring)}
  \label{fig:always_out}
\end{subfigure}%
\hspace{5em}
\begin{subfigure}[t]{.35\textwidth}
  \centering
  \includegraphics[width=.85\linewidth]{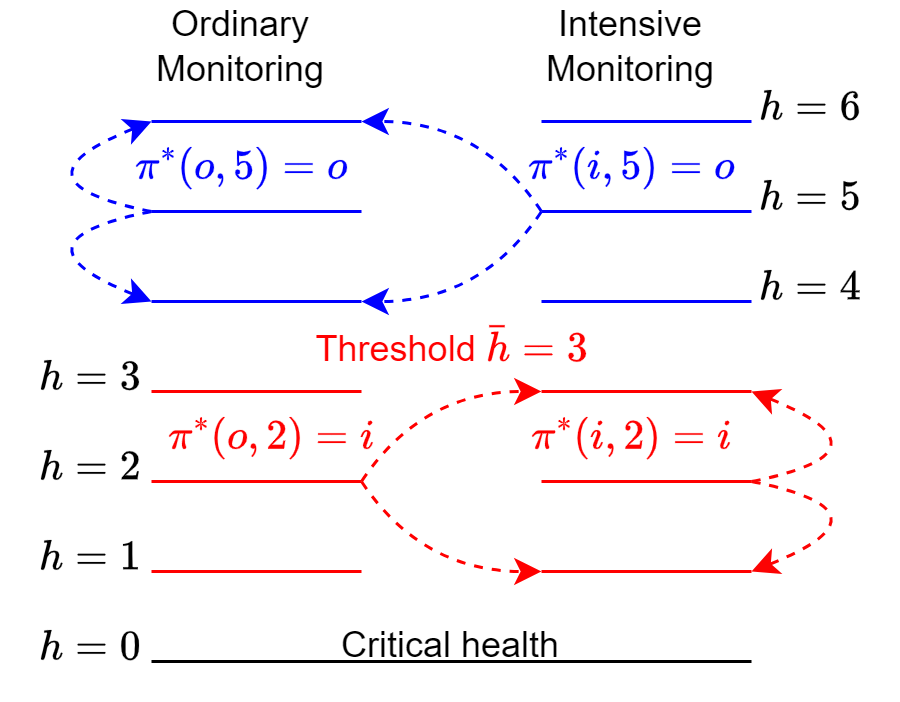}
  \caption{Optimal policy is $\pithresh$ (threshold policy with $\bar{h}=3$)}
  \label{fig:threshold}
\end{subfigure}
\caption{Optimal policies (numerically computed) for $H=6$ and two different parameter sets. (a) For  $\lambda_o=0.2, \lambda_i=0.3,C_c=20, C_i=1, C_o=0, \gamma=0.9$ the optimal policy is $\pi_o$ (using ordinary monitoring at all health states). (b) For $\lambda_o=0.2, \lambda_i=0.3,C_c=60, C_i=1, C_o=0, \gamma=0.9$ the optimal policy is $\pithresh$ with $\bar{h}=3$, that is, ordinary monitoring is used for $h>3$ and intensive for $h\le3$.  }
\label{fig:policies}
\end{figure*} 

An important implication of the above lemma is that the policy where the patient chooses to stay in intensive monitoring for all health states is never optimal. Our next theorem shows that the policy $\pi_o$ is actually optimal for a large choice of parameters.  
\begin{theorem}\label{thm:out}
    Under Assumption 2, the policy $\pi_o$ is optimal ($\pistar=\pi_o$) for the simplified RPM (Definition 1) when the parameters satisfy $$\gamma(\lambda_i-\lambda_o)(1-\phi^2)\leq \frac{C_i}{C_c}.$$
\end{theorem}
\begin{proof}
    See Appendix.\ \ref{app:proofs} for proofs.
\end{proof}

We next define a threshold-policy $\pithresh$ characterized by the health state $\hthresh>0$. These are the policies under which there exists a threshold $\hthresh$ such that the patient stays in intensive monitoring when the patient's health is below or at the threshold $\hthresh$ and in ordinary monitoring when their health is better than $\hthresh$. Note that $\pithresh\in\Pitilde$. So $\pithresh(h)=i$ for $1\leq h\leq \hthresh$ and $\pithresh(h)=o$ for $h>\hthresh$. Our next theorem gives a set of conditions under which $\pithresh$ is the optimal policy for some threshold $\hthresh$.
\begin{theorem}\label{thm:threshold}
    Under Assumption 2, the policy $\pithresh$ is optimal ($\pistar=\pithresh$) for some threshold $\hthresh$ for the simplified RPM (Definition 1) when the following two conditions are satisfied:
    \begin{enumerate}[label=\alph*)]
        \item $\gamma(\lambda_i-\lambda_o)(1-\phi^2)>\frac{C_i}{C_c}$.
        \item $\frac{\gamma\mu_o(1+\gamma\mu_o)}{1-\gamma^2\lambda_o\mu_o}\leq 1$. 
    \end{enumerate}
\end{theorem}
\begin{proof}
    See Appendix.\ \ref{app:proofs} for proofs.
\end{proof}
Condition a) above is the complement of the condition in Theorem \ref{thm:out}. Condition b) is an additional condition which our proof requires for the threshold policy to be optimal. In the asymptotic regime, we strongly believe that condition b) is not necessary, and condition a) alone is sufficient. Hence we believe that in the asymptotic regime, condition a) alone dictates what the optimal policy is and that the optimal policy can only be of two forms - $\pi_o$ and $\pithresh$. This is reinforced by the numerical analysis, presented next.

\section{Performance} \label{sec:perfomance}
In this section, we glean insights on the optimal policy by numerically solving the dynamic programming equations given by \eqref{dp-general}-\eqref{dp-final} to find the optimal policy. 

Figure \ref{fig:policies} depicts the two policies discussed in the last section and a sample set of parameters under which they are optimal. Figure \ref{fig:always_out} shows the policy $\pi_o$, under which the patient stays in ordinary monitoring at all health states. Figure \ref{fig:threshold} shows the policy $\pithresh$ with threshold $\bar{h}=3$, where the patient stays in intensive monitoring for health states $h\leq 3$ and in ordinary monitoring for health states $h>3$. Let model \ref{fig:always_out} use the set of parameters $\lambda_o=0.2, \lambda_i=0.3,C_c=20, C_i=1, C_o=0, \gamma=0.9$ and model \ref{fig:threshold} use the set of parameters $\lambda_o=0.2, \lambda_i=0.3,C_c=60, C_i=1, C_o=0, \gamma=0.9$. Then $\pi_o$ and $\pithresh$ with $\bar{h}=3$ are optimal for model \ref{fig:always_out} and \ref{fig:threshold}, respectively. 

Note that the parameters for model \ref{fig:always_out} satisfy $\gamma(\lambda_i-\lambda_o)(1-\phi^2)\leq C_i/C_c$, which is sufficient for Theorem \ref{thm:out} to hold. Similarly the parameters for model \ref{fig:threshold} satisfy $\gamma(\lambda_i-\lambda_o)(1-\phi^2)> C_i/C_c$, which is condition (a) in Theorem \ref{thm:threshold}. Note, however, that the parameters in model \ref{fig:threshold} do not satisfy condition (b) of Theorem \ref{thm:threshold}, implying that the condition is not necessary. 

When $H$ is finite, there also exist instances where the optimal policy is $\pi_i$, where the patient chooses to stay in intensive monitoring for all health states $h$. But we observed that this policy is optimal only in extreme cases where $H$ is very small or $\gamma$ is very close to $1$. Hence we do not further analyse this policy here.

\begin{figure}[h!]
\centering
\includegraphics[width=0.9\linewidth]{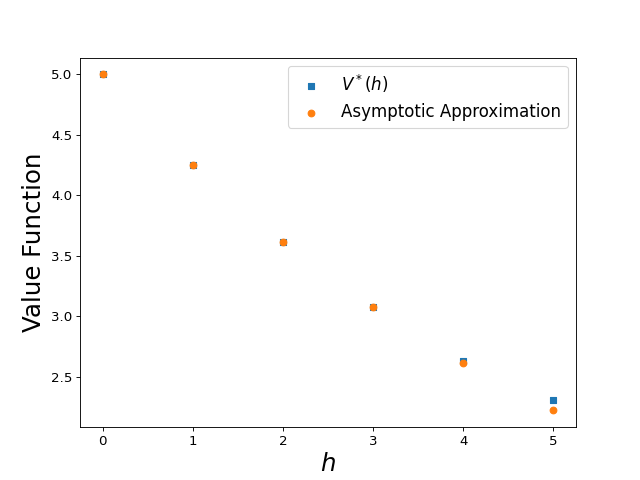}
\caption{The optimal (blue) value function $V^*(h)$ (numerically computed) compared to its asymptotic counterpart (red, $V_o(h)$ obtained from Lemma \ref{lemma:out}) for various health states $h$ in a system with $H=5, \lambda_o=0.2, \lambda_i=0.4,C_c=5, C_i=1, C_o=0, \gamma=0.9$. Note the close proximity of the two plots.}
\label{fig:asymp}
\end{figure}

\begin{figure*}[h!]
\centering
\begin{subfigure}[t]{0.4\textwidth}
  \centering
  \includegraphics[width=0.9\linewidth]{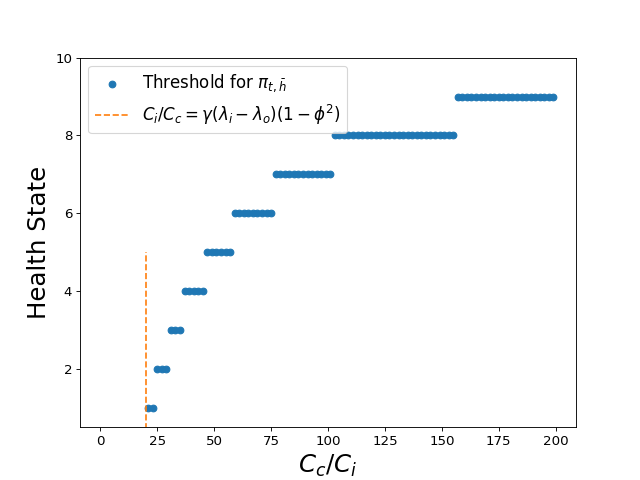}
  \caption{Variation in optimal policy with $C_c/C_i$}
  \label{fig:var_cost}
\end{subfigure}%
\begin{subfigure}[t]{0.4\textwidth}
  \centering
  \includegraphics[width=.9\linewidth]{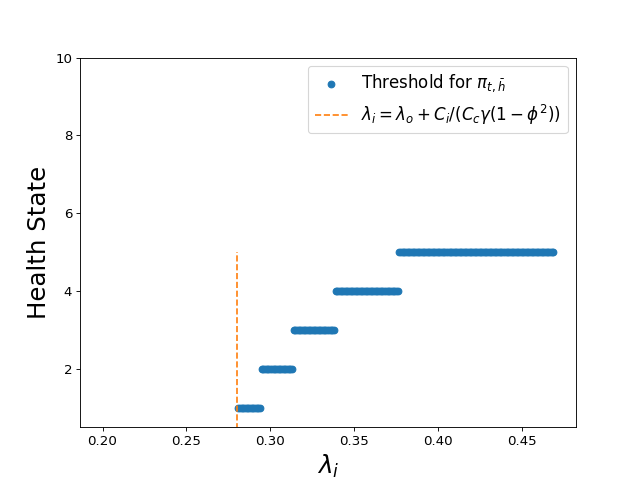}
  \caption{Variation in optimal policy with $\lambda_i$}
  \label{fig:var_lambda}
\end{subfigure}
\begin{subfigure}[t]{.4\textwidth}
  \centering
  \includegraphics[width=.9\linewidth]{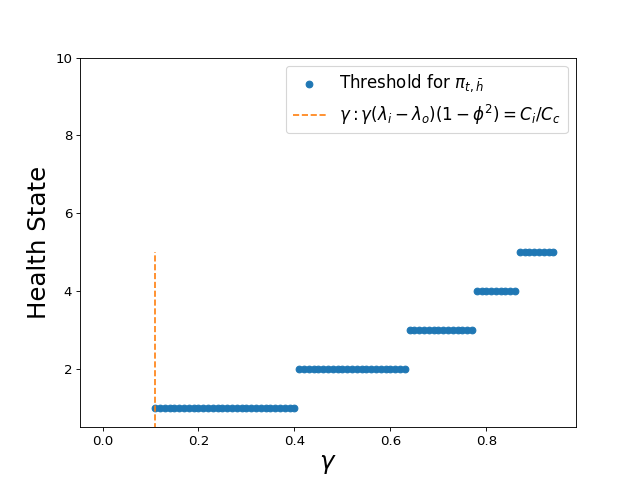}
  \caption{Variation in optimal policy with $\gamma$}
  \label{fig:var_gamma}
\end{subfigure}
\caption{Dependence of the (numerically computed) optimal monitoring policy  on the (a)  cost ratio $C_c/C_i$ (with fixed $\lambda_o=0.2, \lambda_i=0.4, C_i=1, C_o=0, \gamma=0.9$); on (b) $\lambda_i$
(with fixed $\lambda_o=0.2, C_c=50, C_i=1, C_o=0, \gamma=0.9$); and on (c) $\gamma$ (with fixed $\lambda_o=0.2,\lambda_i=0.4, C_c=50, C_i=1, C_o=0$). We set $H=10$ in all cases. Below the vertical orange dashed line, the optimal policy is $\pi_o$ (ordinary monitoring is used for all health states). This line is positioned at the point where the condition of Theorem \ref{thm:out} achieves equality. Above this value the policy changes to $\pithresh$ and each threshold $\bar{h}$ is marked in blue.} 
\label{fig:variation}
\end{figure*} 

Figure \ref{fig:asymp} shows how closely our asymptotic analysis in Section \ref{sec:analysis} relates to the actual solution of the dynamic programming equations. The  parameters are chosen such that the optimal policy is $\pi_o$. We calculate $V^*(h)$ for this model, and compare it with the value function $V_o(h)=\phi^hC_c$ obtained for the asymptotic case $H\to\infty$ (Lemma \ref{lemma:out}). As observed in the plot, the value function obtained for $H=5$ is almost identical to the asymptotic approximation $V_o(h)$. The accuracy of the asymptotic approximation in predicting the optimal policy is further demonstrated by our next result. 

We next study the impact of different parameters on the optimal policy $\pistar$. Figure \ref{fig:var_cost} shows how the optimal policy (numerically computed) varies with the cost ratio $C_c/C_i$. For $C_c/C_i<20.02$ the optimal policy is $\pi_o$, while for $C_c/C_i>20.02$ it is $\pithresh$ with varied thresholds. Note that  $C_c/C_i=20.02$ satisfies the condition in Theorem \ref{thm:out} with equality (the value of $C_c/C_i$ which satisfies $\gamma(\lambda_i-\lambda_o)(1-\phi^2)= C_i/C_C$). This shows that the condition obtained under the asymptotic assumption is a good indicator for our original problem with finite health states. As the ratio $C_c/C_i$ grows, the cost incurred on reaching the critical health state increases, and it gets optimal for the patient to stay under intensive monitoring till their health significantly improves.

Figure \ref{fig:var_lambda} shows how the optimal policy varies as the probability $\lambda_i$ increases. The optimal policy is $\pi_o$ for $\lambda_i<0.28$ and $\pithresh$ with varied thresholds for $\lambda_i>0.28$. Again, $\lambda_i=0.28$ solves the condition in Theorem \ref{thm:out} with equality. As the probability $\lambda_i$ increases, the probability of the patient's health improving under intensive monitoring improves, incentivizing the patient to stay under intensive monitoring for longer. Finally, Figure \ref{fig:var_gamma} shows the impact of $\gamma$ on the optimal policy. As $\gamma$ increases, the patient incurs a higher discounted cost on reaching the critical state, and hence they stay under intensive monitoring for longer.

\section{Conclusions and Extensions}\label{sec:conclusions}

We have developed a two-tier service architecture for remote patient  monitoring (RPM), where the service policy decides whether to place the patient under ordinary or intensive monitoring, given their health state. The optimal policy is first analyzed in asymptotic regimes and conditions are established for choosing ordinary vs intensive monitoring. The policy is then numerically computed and the dependence of its behavior on various key parameters is investigated.

An important extension would be to consider a more general model, which includes non-zero transition costs. Based on numerical experiments performed in the general case, the optimal policy in this case would be a threshold policy with two thresholds instead of the one observed in this paper. A patient under ordinary monitoring would be switched to intensive when their health state deteriorates below a certain lower health threshold, and a patient under intensive monitoring would switch to ordinary when their health state improves above an upper health threshold. There are also other direct extensions, e.g., the costs and probabilities of transitions could also be made dependent on the health state, allowing for a more realistic model.

% \bibliographystyle{IEEEtran}
% \bibliography{refs}


\appendices

\section{Proofs}\label{app:proofs}

\subsection{Proof for Lemma \ref{lemma:out}}

\begin{proof}
    \begin{enumerate}[label=\alph*)]
        \item For the policy $\pi_o$, the value function is $$V_o(h)=\EE\left[\sum_{t=0}^{T-1}(\gamma^tc(s_t,\pi_o(s_t)))+\gamma^TC_c|h_0=h\right],$$ for all $h\geq1$. Here $T$ denotes the time at which the patient reaches health state $0$. Since $\pi_o(s_t)=o$, $c(s_t,\pi_o(s_t))=0$ for all $t<T$. This implies that $V_o(h)=C_c\EE[\gamma^T|h_0=h]$, where $T$ is the time at which the patient reaches health state $0$. 

        Consider a infinite state 1-dimensional random walk with probability of moving forward $\mu_o$ and probability of moving backward $1-\mu_o=\lambda_o$. Let $t_{0,h}$ be the time taken to hit state $h$ for a random walk initialized at state $0$. Then $T$ defined above is equal to $t_{0,h}$ for this random walk and $\EE[\gamma^T|h_0=h]=\EE[\gamma^{t_{0,h}}]$. Note that $t_{0,h}=t_{0,1}+t_{1,2}+\ldots,t_{h-1,h}$. Now,
        \begin{align*}
        \EE[\gamma^{t_{0,h}}]&=\EE[\gamma^{t_{0,1}+t_{1,2}+\ldots t_{h-1,h}}]\\
        &=\EE[\gamma^{t_{0,1}}\times\gamma^{t_{1,2}}\times\ldots\times\gamma^{t_{h-1,h}}] \\
        &\stackrel{(a)}{=}\EE[\gamma^{t_{0,1}}]\times \EE[\gamma^{t_{1,2}}]\times\ldots\times\EE[\gamma^{t_{h-1,h}}]\\
        &\stackrel{(b)}{=}\EE[\gamma^{t_{0,1}}]^h
        \end{align*}
        Here equality (a) follows from the independence of $t_{0,1},\ldots,t_{h-1,h}$ and equality (b) follows from the fact that $t_{0,1},\ldots,t_{h-1,h}$ follow the same distribution.
 Since $\mu_o>\lambda_o$, the probability that the patient reaches state $1$ in the random walk starting at state $0$ is $1$. Then \cite[Chapter XIV, eqn.\ 4.8]{prob-book} gives us $\EE[\gamma^{t_{0,1}}]=\phi$ and 
 $V_o(h)=C_c \phi^h,$
        where $$\phi=\frac{1-\sqrt{1-4\lambda_o\mu_o\gamma^2}}{2\lambda_o\gamma}.$$
        \item Let $h'=\lceil\log(C_i/C_c)/\log(\phi)\rceil+1$. Then $$V_o(h')= C_c\phi^{h'}\leq C_c\phi\times (C_i/C_c)=\phi C_i.$$ Since $\phi\leq 1$, note that 
        $V_o(h)\leq V_o(h')\leq \phi C_i$ for all $h\geq h'$. Consider a policy $\pi'$ such that $\pi'(h)=i$. Then $V_{\pi'}(h)\geq C_i$. Now for any $h\geq h'$, $V_o(h)<V_{\pi'}(h)$ as $\phi<1$ for $\gamma<1$. This implies that the policy $\pi'$ cannot be optimal, and hence under the optimal policy, the patient prefers to stay in ordinary monitoring for all health states $h\geq h'$. This completes the proof for Lemma \ref{lemma:out}.
    \end{enumerate}
\end{proof}
\subsection{Proof for Theorem \ref{thm:out}}
\begin{proof}
We know that a policy $\pi$ is optimal if and only if
\begin{align*}
    &c(s,\pi(s))+\gamma\sum_{s'}p(s'|s,\pi(s))V_\pi(s')\\
    &=\min_{a}\left(c(s,a)+\gamma\sum_{s'}p(s'|s,a)V_\pi(s')\right),
\end{align*}
is true for all states $s$ \cite[Proposition 2.2 and 2.3]{Bertsekas}. Hence policy $\pi_o$ is optimal if and only if  
\begin{align*}
&c(s,o)+\gamma\sum_{s'}p(s'|s,o)V_o(s')\\
    &\leq c(s,i)+\gamma\sum_{s'}p(s'|s,i)V_o(s'),
\end{align*}
for all states $s$. This implies that $\pi_o$ is optimal if and only if  
\begin{align}\label{proof-thm1-eqn}
    &\gamma(\lambda_o V_o(h+1)+\mu_oV_o(h-1))\nonumber\\
    &\leq C_i+\gamma(\lambda_iV_o(h+1)+\mu_iV_o(h-1)),
\end{align}
for all $1\leq h$. Subsituting value of $V_o(\cdot)$ from Lemma \ref{lemma:out}, we have that $\pi_o$ is optimal if and only if
\begin{align*}
    & C_c\gamma(\lambda_o\phi^{h+1}+\mu_o\phi^{h-1})\leq C_i+C_c\gamma (\lambda_i\phi^{h+1}+\mu_i\phi^{h-1})\\
    &\iff C_c\gamma \phi^{h-1}((\lambda_o-\lambda_i)\phi^2+(\mu_o-\mu_i))\leq C_i\\
    &\stackrel{(b)}{\iff} C_c\gamma \phi^{h-1}(\lambda_i-\lambda_o)(1-\phi^2)\leq C_i,
\end{align*}
for all $1\leq h$. Here (b) follows from the definition that $\mu_o=1-\lambda_o$ and $\mu_i=1-\lambda_i$. 

Note that 
\begin{align*}
    &\gamma(\lambda_i-\lambda_o)(1-\phi^2)\leq \frac{C_i}{C_c}\\
    &\implies C_c\gamma (\lambda_i-\lambda_o)(1-\phi^2)\leq C_i\\
    &\implies C_c\gamma \phi^{h-1}(\lambda_i-\lambda_o)(1-\phi^2)\leq C_i\;\;\forall h\geq 1.\\
    &\implies \gamma(\lambda_o V_o(h+1)+\mu_oV_o(h-1))\nonumber\\
    &\;\;\;\;\;\;\;\;\;\;\;\leq C_i+\gamma(\lambda_iV_o(h+1)+\mu_iV_o(h-1))\;\;\forall h\geq 1.
\end{align*}
Hence if $\gamma(\lambda_i-\lambda_o)(1-\phi^2)\leq \frac{C_i}{C_c}$ is satisfied then the policy $\pi_o$ is optimal. This completes the proof for Theorem \ref{thm:out}.
\end{proof}
\subsection{Proof for Theorem \ref{thm:threshold}}
\begin{proof}
    Since $\gamma(\lambda_i-\lambda_o)(1-\phi^2)>\frac{C_i}{C_c}$, eqn.\ \eqref{proof-thm1-eqn} is not satisfied for $h=1$. This implies that $\pi_o$ is not optimal. This implies that under the optimal policy the patient will stay under intensive monitoring for some state. Define $Q^*(h,i)=C_i+\gamma(\lambda_i\Vstar(h{+}1)+\mu_i\Vstar(h{-}1))$ (respectively, $Q^*(h,o)=\gamma(\lambda_o\Vstar(h{+}1)+\mu_o\Vstar(h{-}1))$). $Q^*(h,i)$ denotes the Q-functions, where action $i$ (or $o$, respectively) is taken when initialized at health state $h$ and then actions are taken using the optimal policy. Note that $\pistar(h)=o$ if $Q^*(h,o)<Q^*(h,i)$, and $\pistar(h)=i$ otherwise.
    
    Suppose $Q^*(h,o)-Q^*(h,i)$ is monotonically decreasing as $h$ increases, then if action $o$ is optimal at some $h$,     
    then it will also be optimal at $h+1$ and so on (i.e., $Q^*(h,o)-Q^*(h,i)\leq 0\implies Q^*(h+1,o)-Q^*(h+1,i)$). As condition (a) already enforces that policy $\pi_o$ is not optimal, action $i$ has to be taken at state under the optimal policy. Also, Lemma \ref{lemma:out} part (b) shows that there exists $h'$ such that the optimal policy for health states above $h'$ is $o$, which implies that policy $o$ is played at some state. Hence    
    the optimal policy has to be $\pithresh$ for some $\bar{h}$.

    Hence we just need to show that $Q^*(h,o)-Q^*(h,i)$ is monotonically decreasing as $h$ increases to prove that $\pithresh$ is optimal.  
    
    We can show that $$Q^*(h,o)-Q^*(h,i)=\gamma(\lambda_i-\lambda_o)(\Vstar(h-1)-\Vstar(h+1))-C_i.$$ Now, 
    \begin{align*}
        &\big(Q^*(h+1,o)-Q^*(h+1,i)\big)-\big(Q^*(h,o)-Q^*(h,i)\big)\\
        &=\gamma(\lambda_i-\lambda_o)\Big(\Vstar(h)-\Vstar(h+2)\\
        &\;\;\;\;\;\;\;\;\;\;\;\;\;\;\;\;\;\;\;\;\;\;\;\;-\Vstar(h-1)+\Vstar(h+1)\Big)
    \end{align*}
    Hence if $\Vstar(h)+\Vstar(h{+}1)\leq \Vstar(h{-}1)+\Vstar(h{+}2)$ is true for all $h\geq 1$ then $Q^*(h,o){-}Q^*(h,i)$ is monotonically decreasing with $h$. Now we know using \cite[Proposition 2.2]{Bertsekas} that 
    $$\Vstar(h)\leq \gamma\lambda_o\Vstar(h+1)+\gamma\mu_o\Vstar(h-1),$$
    and 
    $$\Vstar(h+1)\leq \gamma\lambda_o\Vstar(h+2)+\gamma\mu_o\Vstar(h).$$ 
    With some further manipulation, we can show that 
    \begin{align*}
        \Vstar(h)+\Vstar(h+1)&\leq\frac{\gamma\lambda_o(1+\gamma\lambda_o)}{1-\gamma^2\lambda_o\mu_o}\Vstar(h+2)\\
    &+\frac{\gamma\mu_o(1+\gamma\mu_o)}{1-\gamma^2\lambda_o\mu_o}\Vstar(h-1).
    \end{align*}
    We can show that $\frac{\gamma\lambda_o(1+\gamma\lambda_o)}{1-\gamma^2\lambda_o\mu_o}$ is always less than $1$ for $\lambda_o\leq 0.5$. Hence if $\frac{\gamma\mu_o(1+\gamma\mu_o)}{1-\gamma^2\lambda_o\mu_o}\leq 1$, then $\Vstar(h)+\Vstar(h{+}1)\leq \Vstar(h{-}1)+\Vstar(h{+}2)$ and hence $Q^*(h,o){-}Q^*(h,i)$ is monotonically decreasing with $h$. Under the additional condition that $\gamma(\lambda_i-\lambda_o)(1-\phi^2)>\frac{C_i}{C_c}$, this implies that $\pithresh$ is the optimal policy for some threshold $\bar{h}$.
\end{proof}

\end{document}